\documentclass[lettersize,journal]{IEEEtran}
\usepackage{graphicx}
\usepackage{subfigure}
\usepackage{float}
\usepackage{mathrsfs}
\usepackage{amsfonts}
\usepackage{array}
\usepackage{amssymb,amsmath}
\usepackage{longtable}
\usepackage{rotating}
\usepackage{multirow}
\usepackage{epstopdf}
\usepackage[lined,boxed,commentsnumbered, ruled]{algorithm2e}
\usepackage{cite}
\usepackage{float}
\usepackage{stfloats}
\usepackage{algorithmic}
\usepackage[colorlinks,
            linkcolor=black,
            anchorcolor=black,
            urlcolor=blue,
            citecolor=black]{hyperref}
\usepackage[hyphenbreaks]{breakurl}

\renewcommand{\baselinestretch}{1}

\def\BibTeX{{\rm B\kern-.05em{\sc i\kern-.025em b}\kern-.08em
    T\kern-.1667em\lower.7ex\hbox{E}\kern-.125emX}}

\renewcommand{\baselinestretch}{1}
\newcommand{\systemname}{{{CA-FedRC}}}
\usepackage{color}

\begin{document}
\title{CA-FedRC: Codebook Adaptation via Federated Reservoir Computing in 5G NR}
\author{
Ziqiang Ye,\IEEEmembership{~}
Sikai Liao,\IEEEmembership{~}
Yulan Gao, \IEEEmembership{Member, IEEE,}
Shu Fang, \IEEEmembership{Member, IEEE,}
Yue Xiao, \IEEEmembership{Member, IEEE,} \\
Ming Xiao, \IEEEmembership{Senior Member, IEEE}
and Saviour Zammit, \IEEEmembership{Senior Member, IEEE}
\thanks{Z. Ye, S. Liao, S. Fang, and Y. Xiao are with the National Key Laboratory of Wireless Communications, University of Electronic Science and Technology of China, Chengdu 611731, China (e-mail: yysxiaoyu@hotmail.com, sikailiao@163.com, fangshu@uestc.edu.cn, xiaoyue@uestc.edu.cn).}
\thanks{Y. Gao and M. Xiao are with the Department of Information Science and Engineering, Royal Institute of Technology (KTH), 10044 Stockholm, Sweden (e-mail: yulang@kth.se, mingx@kth.se).}
\thanks{S. Zammit is with the Department of Communications and Computer Engineering, University of Malta, Msida MSD 2080, Malta (e-mail: saviour.zammit@um.edu.mt).}
\vspace{-4mm}
}

\maketitle

\begin{abstract} 
With the burgeon deployment of the fifth-generation new radio (5G NR) networks, the codebook plays a crucial role in enabling the base station (BS) to acquire the channel state information (CSI).
Different 5G NR codebooks incur varying overheads and exhibit performance disparities under diverse channel conditions, necessitating codebook adaptation based on channel conditions to reduce feedback overhead while enhancing performance.
However, existing methods of 5G NR codebooks adaptation require significant overhead for model training and feedback or fall short in performance.
To address these limitations, this letter introduces a federated reservoir computing framework designed for efficient codebook adaptation in computationally and feedback resource-constrained mobile devices.
This framework utilizes a novel series of indicators as input training data, striking an effective balance between performance and feedback overhead. 
Compared to conventional models, the proposed codebook adaptation via federated reservoir computing (\systemname{}), achieves rapid convergence and significant loss reduction in both speed and accuracy. 
Extensive simulations under various channel conditions demonstrate that our algorithm not only reduces resource consumption of users but also accurately identifies channel types,  
thereby optimizing the trade-off between spectrum efficiency, computational complexity, and feedback overhead.

\end{abstract}


\section{Introduction}

In the fifth-generation new radio multiple-input multiple-output (5G NR MIMO) system, the base station (BS) leverages downlink channel state information (CSI) for precoding to boost the spectral efficiency of the physical downlink shared channel (PDSCH) \cite{dahlman20205g}.
The acquisition of CSI is a critical research area in 5G NR MIMO systems due to the computational and feedback constraints commonly associated with user equipment (UE) \cite{hussain2020limitations}. 
Consequently, a codebook-based feedback approach is widely adopted in practical implementations.
Ref. \cite{38.214} defined coarse resolution codebook with less feedback overhead as release Type I and fine resolution codebook with better performance but heavier feedback overhead as release enhanced Type II (EType II).
However, the performance of Etype II significantly deteriorates in scenarios characterized by high spatial correlation, frequency selectivity and high mobility.
In such cases, it may even be outperformed by the Type I codebook, but lead to increased power consumption at the UE due to extensive calculations and feedback overhead \cite{qin2023review}. 

Therefore, it is critical to monitor and identify channel types and scenarios where the EType II codebook underperforms, then adaptively switch to the Type I codebook to reduce the computational complexity and feedback overhead.
For example, Ref. \cite{10077214} detailed methods using random forest (RF), K-nearest-neighbor (KNN), and support vector machine (SVM) techniques to classify channels based on mobility, delay profile, and line-of-sight (LoS) characteristics, establishing a foundation for codebook adaptation strategies between Type I and EType II.
Additionally, the authors in \cite{8993100} ultilized convolutional neural networks (CNNs) to classify non-line-of-sight (NLoS) and LoS channels, although this relies on extensive datasets from UE channel measurements, which is impractical for UEs with limited computational capacity, particularly in 5G NR systems with numerous antennas and broad frequency resources.
In \cite{US}, the channel characteristics detrimental to EType II were discussed, with traditional logic used to determine the codebook type.
However, obtaining precise decision thresholds for this approach remains challenging.

The emergence of reservoir computing (RC) \cite{lukovsevivcius2009reservoir,jaeger2007echo} has provided an unparalleled tool in the field, recognized for its impressive predictive performance and the effective trade-off it offers between accuracy and efficiency in learning.
However, it should be noted that the echo state networks (ESNs), a subset of RC, may sometimes lead to suboptimal performance \cite{zheng2021enhancing} due to their inherent simplicity.
The advent of federated learning (FL) has addressed some of these concerns by allowing users to collaboratively refine their models while maintaining low communication overhead.
Notably, CSI is typically transmitted unencrypted on control-plane physical channels, emerges as a valuable source for localization \cite{roth2021localization}, necessitating robust measures to protect CSI from potential leakage.
FL enhances privacy by transmitting only model gradients, rather than raw data, thereby not only improving the model's generalization capabilities across various user datasets \cite{konevcny2016federated}, but also ensuring formal privacy protections.
Given these advancements, integrating FL with reservoir computing represents a significant step forward compared to traditional methods for tackling the challenges previously discussed.

Taking these considerations into account, this letter introduces a novel codebook adaptation approach designed to  balance the spectral efficiency with feedback overhead tax.
By optimizing the utility, which is influenced by the weighted spectrum efficiency gain and feedback overhead tax, we address the conflict between the high demand of communication capacity and the overhead limitation of UEs.
To tackle this balancing problem, we introduce a federated reservoir computing framework, which combines the strength of both federated learning paradigm and reservoir computing.

The innovations and contributions of this letter are outlined as follows:
\begin{itemize}
\item A novel integration of reservoir computing within a federated learning framework for codebook adaption as \systemname{} is constructed, which provides a resource-efficient machine learning approach for PDSCH precoding at BS to dynamically change the codebooks for 5G NR system under different channel types.
\item The proposed \systemname{} realizes the trade-off  between spectrum efficiency and feedback overhead by codebook adaptation based on federated learning of the designed CSI indicators as input training data, which relieves the CSI mismatch problem for PDSCH precoding in 5G NR system, and improves its overall communication performance with significantly reduced feedback overhead.
\item A favorable balance between performance and feedback overhead is achieved, providing approximately 1.6\% average spectral efficiency gain with reduced feedback overhead of 250\%
compared to the EType II benchmark within a broad SNR region.
\end{itemize}


\section{System Model}
\begin{figure*}[!t]
\centering
\includegraphics[width=0.75\textwidth]{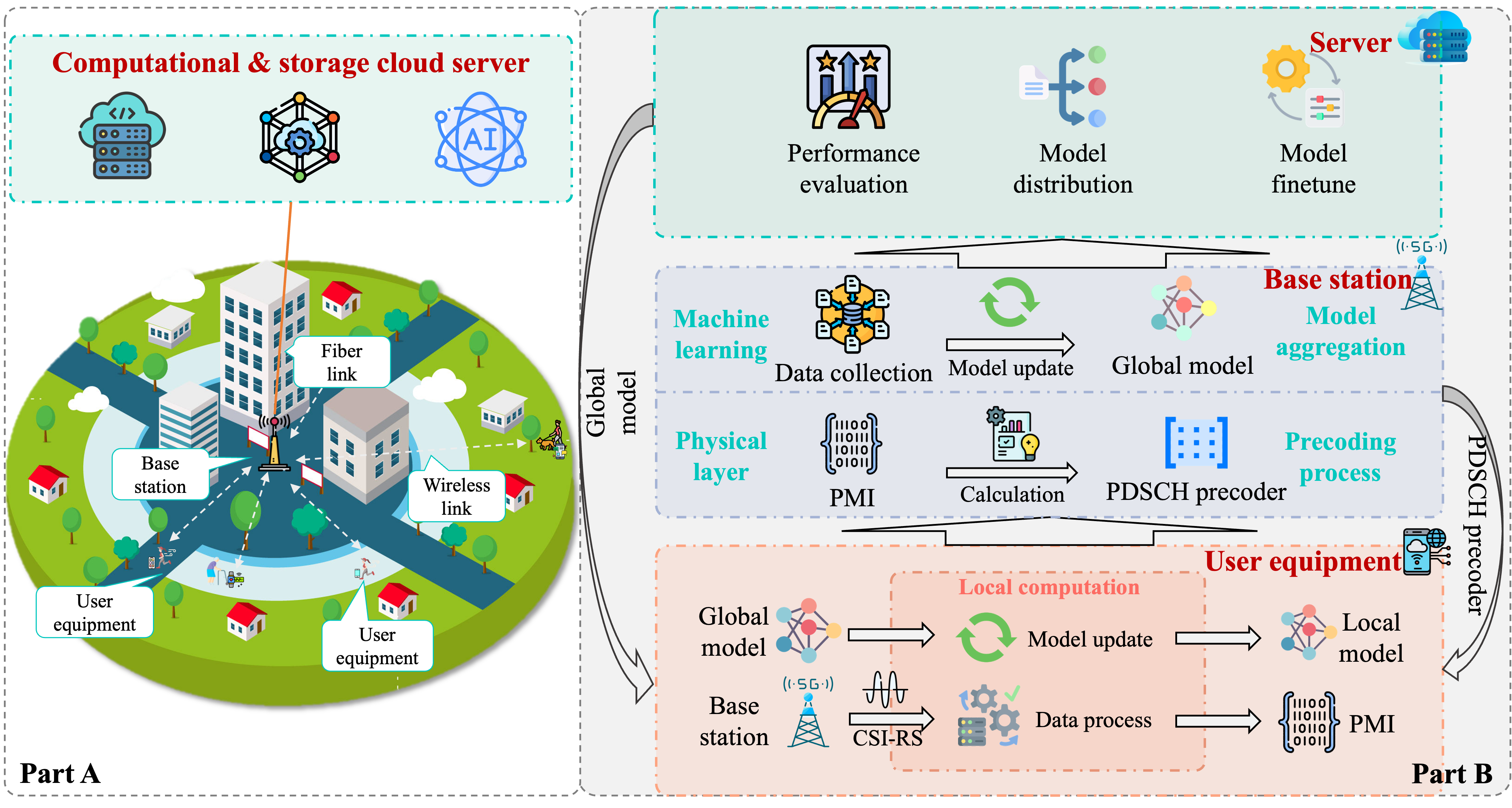}
\caption{The architecture of \systemname{} in cellular system. }
\vspace{-5mm}
\label{fig:1}
\end{figure*}

We consider a general cellular communication system comprising a set ${\mathcal G}$ UEs located within the coverage area of a BS, shown in Fig. \ref{fig:1} (Part A) . 
The BS is employed with $N_{\text{bs}}$ antennas, where each UE within the network is equipped with $N_{\text{ue}}$ antennas, respectively. 
To ensure that any UE can contribute to FL, we assume deploying RC model on each UE.
The entire workflow of \systemname{} is depicted in Fig. \ref{fig:1} (Part B). 
Specifically, each UE employs the channel state information reference signal (CSI-RS) received from the BS to compute  the CSI indicators, which then is served as input data for locally training the RC model at UE.
After local training, each UE selects an optimal codebook based on the output of the model and feeds back the codebook index to the server at BS.
At last, the server outputs the PDSCH precoding matrix for BS via the proposed trade-off strategy. 
In general, the ensuing part introduces the preliminaries of the 5G NR codebook. 


Release $15$ NR Type I codebook $\mathbf{W}$ can be defined by the multiplication of a wide-band beam matrix $\mathbf{W} _1$ and a sub-band phase quantization matrix $\mathbf{W}_2$ as
\begin{align}\label{eq:1}
\begin{array}{ll}
     \mathbf{W} &= \mathbf{W}_1\mathbf{W}_2\\ 
     &=\frac{1}{\sqrt{2RN_1N_2}}\times\left[ \begin{matrix}
        w_{0,0} & w_{0,1} & \cdots w_{0,R-1} \\
        w_{1,0} & w_{1,1} & \cdots w_{1,R-1}
    \end{matrix}\right],
\end{array}
\end{align}
where $N_1$, $N_2$ are the numbers of antenna ports in horizontal and vertical respectively, thus the total number of antenna ports is $N_p=2N_1N_2$.
For polarization $d \in \{0,1\}$, where $0$ and $1$ indicate $+45^\circ$ and $-45^\circ$ polarization respectively. 
The weighted liner combination of $L$ discrete Fourier transformation (DFT) beams $w_{d,r}$ with the transmission rank $r \in \{0,\cdots,R-1\}$ are expressed as
\begin{align}\label{eq:2}
   w_{d,r} = b_{i_{1,1}+k_1,i_{1,2}+k_2}\times c_{d,r},
\end{align}
where $b_{i_{1,1}+k_1,i_{1,2}+k_2}$ is the beam selected from a 2-D DFT matrix that is over-sampled by factors $O_1$ and $O_2$ in horizontal and vertical dimensions, with the horizontal index $i_{1,1}$ and the vertical index $i_{1,2}$ for the selected beam.
Meanwhile, $c_{d,r}$ is the quantified phase combination factor, and $k_1$, $k_2$ indicate the offset of the selected beam for different rank related to the index $i_{1,1}$ and $i_{1,2}$ respectively.
Thus the overall feedback overhead stands for the sum of codebook PMI as $\log_2(N_1O_1)$ bits in the horizontal direction and $\log_2(N_2O_2)$ bits in the vertical direction.
Moreover, there are $2$ bits for beam offset indication, where the index of quantified phase combination requires $2$ bits as $r\!=\!1$ or $1$ bit as $r\!\geq \!2$ for each sub-band \cite{38.214}. 

\begin{figure} 
    \centering
    \subfigure[]{
    \begin{minipage}[t]{0.45\linewidth}
    \includegraphics[width=1.8in]{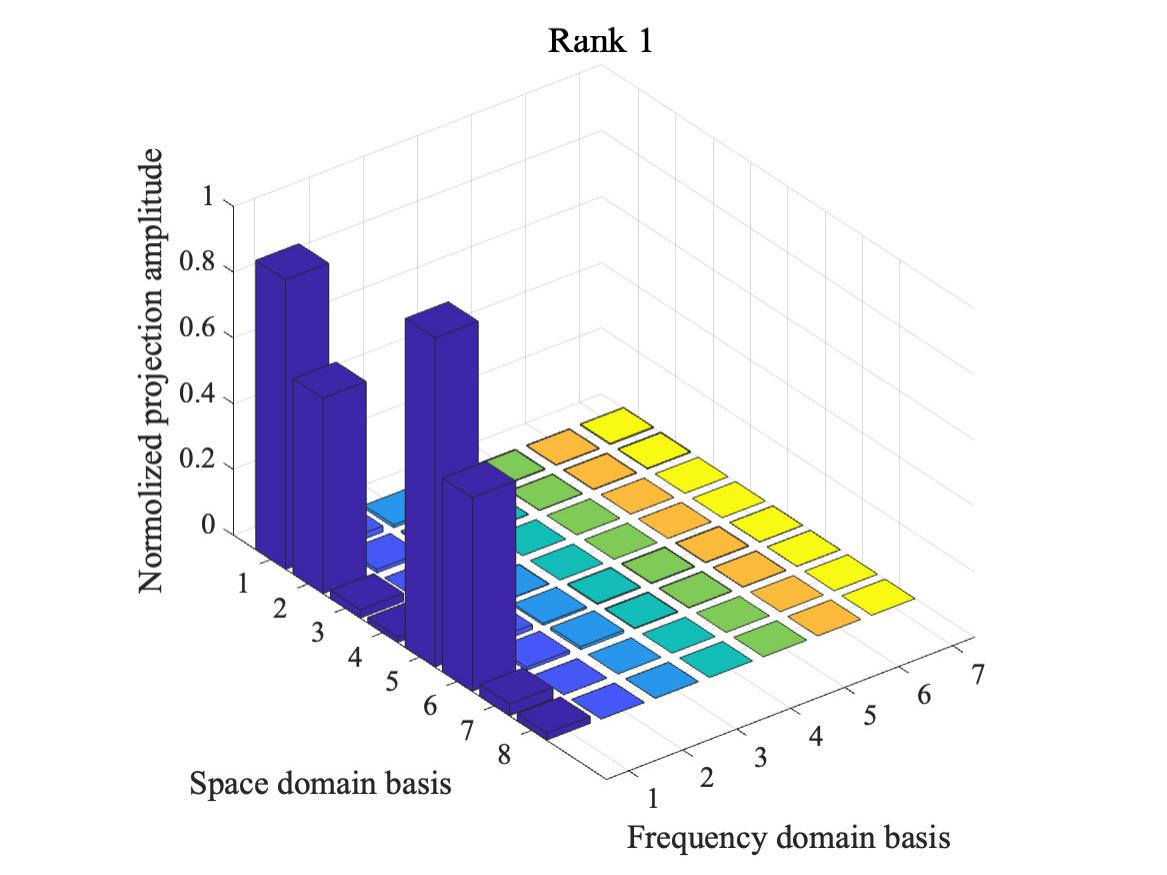}
    \end{minipage}
    \label{fig:3-1}
}
\subfigure[]{
    \begin{minipage}[t]{0.4\linewidth}
    \includegraphics[width=1.8in]{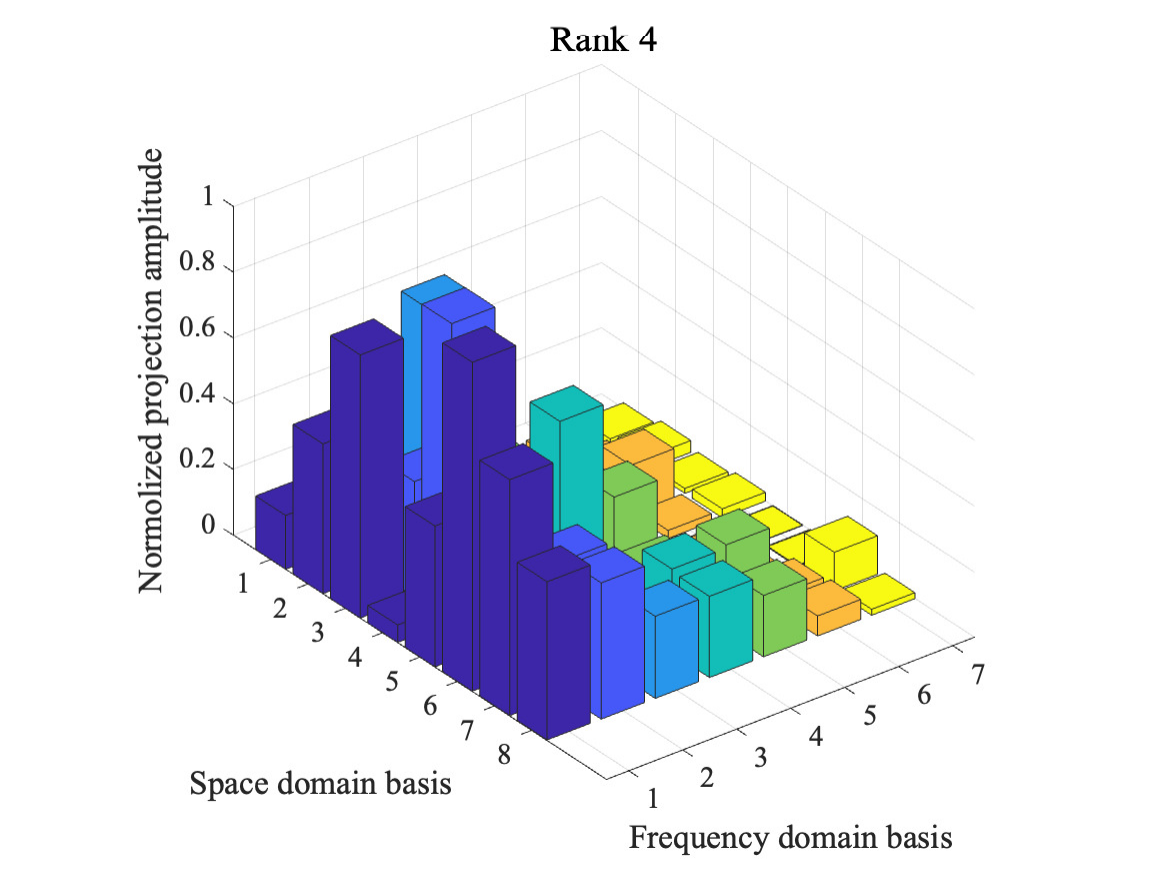}
    \end{minipage}
    \label{fig:3-2}
}
\caption{Normalized projection amplitude of Rank 1 and Rank 4 on frequency and space domain.}
\vspace{-5mm}
\end{figure}

Shown in Eq. \eqref{eq:1}, Type I codebook uses only one DFT vector to form a beam for data transmission in each layer as $R>2$, exhibiting commendable performance in scenarios with high spatial correlation, but yielding severe performance degradation and poor system throughput in rich scattering scenarios.
So EType II codebook $\mathbf{W}^e$ introduces multi-beams to adapt rich scattering environments, which consists of three parts \cite{8727185} as
\begin{align}\label{eq:3}
     \mathbf{W}^e = \mathbf{W}_1\tilde{\mathbf{W}}_2\mathbf{W}_f,
\end{align}
where $\mathbf{W}_1\in \mathbb{C}^{N_p\times 2L}(2L<N_p)$ is the spatial compression matrix with $L$ beams selected from an over-sampled DFT matrix, in which the feedback overhead is measured by the sum of beam-selection index as $\lceil \log_2{\binom{N_1N_2}{L}}\rceil$ bits and over-sampled index as $\lceil \log_2{O_1O_2}\rceil$ bits.
EType II considers compressing the channel in frequency domain with a matrix $\mathbf{W}_f\in \mathbb{C}^{M\times N_{SB}}$ comprised of $M$ DFT bases for $N_{SB}$ sub-bands, whose feedback overhead is the selection index of frequency bases as $\lceil(\log_2{\binom{N_{SB}}{M}}) \rceil$ bits.
Meanwhile, $\tilde{\mathbf{W}}_2 \in \mathbb{C}^{2L\times M}$ is the liner combination matrix with $K_{nz}$ non-zeros coefficients indicating its sparsity due to the high correlation in the space and frequency domains with $5+7K_{nz}+2LM$ bits in each layer.
However, EType II codebook demonstrates a bias towards low-rank transmissions, capitalizing on the sparsity features of the channel as depicted in Fig. \ref{fig:3-1}, while ignoring the rich scattering channel with high rank transmission possibilities as illustrated in Fig. \ref{fig:3-2}, leading to degraded performance and throughput under such scenario.
Another drawback of the EType II codebook is its limited resistance to Doppler effects.
In scenarios involving high-speed mobility, the robustness of the EType II codebook degrades, attributed to its insufficient capability to track rapid changes in channel information.
Moreover, compared with Type I codebook calculated using wide-band information in Eq. \eqref{eq:1}, EType II codebook calculation in Eq. \eqref{eq:3} requires frequency-domain compression to reduce the feedback overhead, that also leads to significant performance loss in high frequency-selective channels. 
Furthermore, under the high spatial correlation channels, EType II codebook consumes much more feedback overhead, but only shows comparable performance to the Type I codebook.
In summary, regardless of high mobility, high spatial correlation or high frequency-selective channels,  EType II codebook suffers from notable performance loss.
Since the high feedback overhead of the EType II codebook not always turns in high performance gain compared with the Type I codebook, a codebook adaptation strategy with federal learning based reservoir computing as \systemname{} is investigated in this contribution for the practical scenarios with dynamic changing channels, towards higher performance and throughput with reduced feedback overhead.

\section{Problem Formulation \& Algorithm Design}
\subsection{Problem Formulation}
To achieve a balanced trade-off between communication performance and codebook feedback overhead $O_{fd}$, an efficient strategy is presented in this section as follows.

Each UE $g$ selects $A_g\in \{0,1\}$ which represents Type I or EType II codebook to maximize its utility as
\begin{align}\label{eq:11}
U(A_g) =\eta_{S\!E}^{gain}(A_g) - \lambda {O^{tax}_{fd}}(A_g),
\end{align}
where $\mathbf{A}=\{A_g|\forall g\in \mathcal{G}\}$.
To obtain the balance between spectral efficiency and feedback overhead, the balance parameter $\lambda \in [0,1]$ is introduced.
The overhead tax, denoted as $O_{fd}^{tax} = O_{fd}(A_g)-O_{fd}^{I}$ is from ETypeII's feedback overhead $O_{fd}^{II}$ substracting Type I's overhead $O_{fd}^{I}$, as detailed in Section II.
The spectral efficiency of Type I codebook $\eta^b_{S\!E}$ is used as the baseline thus the spectral efficiency gain for UE $g$, calculated at UE with the $g$-th UE's adaptive $A_g$, is defined as
\begin{align}\label{eq:13}
   \eta_{S\!E}^{gain}(A_g) = ({\eta_{S\!E}(A_g)}/{\eta^b_{S\!E}}-1),
\end{align}
where $\eta_{S\!E}(A_g)$ represents the overall achievable spectrum efficiency for each UE, calculated with the available spatial and frequency resources.
This is determined by summing the logarithmic function of $1+\text{SINR}_r^n$ across all resource blocks (RBs) within the current CSI-RS period:
$\eta_{S\!E}(A_g)=\sum_{r=1}^{R}\sum_{n=1}^{N_{RB}}\log_2(1+\text{SINR}_r^n)$.
Here, $N_{RB}$ denotes the number of RBs. Each UE computes this locally during training with the coupled channel, then feed backs the CSI-RS training output $A_g$ for each user.
The baseline is $\eta^b_{SE}\!=\! \eta_{S\!E}(A_g\!=\!0)$, and
$\text{SINR}_r^n$ is the signal to interference and noise ratio (SINR) for the $n$-th RB and $r$-th rank for each user, calculated as 
\begin{align}\label{eq:8}
\text{SINR}_r^n = \frac{\vert \mathbf{G}_r^n\mathbf{H}^n\mathbf{F}^n_r \vert ^2}{\sum_{j\neq r}\vert \mathbf{G}_r^n\mathbf{H}^n\mathbf{F}^n_j \vert ^2 +\Vert \mathbf{G}_r^n\Vert_2^2 \sigma_n^2},\\ \notag
\forall r\in\{1,...,R\},\forall n\in \{1,...,N_{RB}\},
\end{align}
where $\mathbf{G}_r^n$ is the equalizer for the $r$-th rank and $n$-th RB, $\sigma_n^2$ is noise variance for each UE of the $n$-th RB, and
$\mathbf{F}^n_r$ is the precoding matrix for PDSCH of the $n$-th RB, defined as the multiplication of precoding matrix for CSI-RS training pilot $\mathbf{W}^{csi}$ and feedback codebook matrix $\mathbf{W}^n_r$ for the $r$-th rank and $n$-th RB.
Thus we have
\begin{align}\label{eq:12}
   \mathbf{F}_r^n= \mathbf{W}^{csi}\mathbf{W}_r^n.
\end{align}

Therefore, $\eta_{S\!E}$ represents the ideal and predicted achievable spectrum efficiency, which cannot be realized in practical system. In reality, systems can only utilize feedback from previous CSI-RS to inform the transmission strategies for the current channel.
This limitation reflects the inherent delay and approximation involved in applying historical channel data to present conditions.
This problem is regarded as CSI mismatch for precoding techniques, especially under high mobility scenarios.
For the above process, UE $g$ obtains the adaptive $A_g$ for maximizing the utility over the current CSI-RS period according to Eqs. \eqref{eq:11}-\eqref{eq:12} under current CSI-RS and local model $\Omega_g$ to alleviate the CSI mismatch problem. $\Omega_g$ is the local model with input layer, a randomly connected dynamic reservoir and a trainable output layer with a Sigmoid activation function.
That is also one of the motivations of the proposed method with \systemname{} for tracking the optimal codebook with the the moving terminals.
\vspace{-5mm}
\subsection{\systemname{} Realization}

The proposed codebook adaptation via federated reservoir computing, namely \systemname{}, is presented in this part with a novel FL approach \cite{mcmahan2017communication} that incorporates RC with codebook adaptation for enhancing the 5G NR system. 
The algorithm commences by initializing a global ESN model, an efficient ML approach and a local ESN model for each UE, as Fig. \ref{fig:1} (Part B) shows.
Specifically, the architecture of an ESN consists of two components, where one is the recurrent network that holds an internal state over the time steps, called the reservoir.
The other component, the readout, is a linear layer that takes as input state of the reservoir and outputs a prediction  by training.
Due to the nature of reservoir computing, training process only happens in the readout layer, which allows for extremely efficient training process.
By introducing FL in the RC, the proposed \systemname{} guarantees the confidentiality of the CSI-RS and reduces the communication overhead as the UE uploads the gradient of the local model rather than the original data.

In the training stage, UE $g$ obtains the local model from the server and constructs the local dataset $\mathcal{X}_g$ by CSI indicator data $\mathbf{X}_g \in \mathcal{X}_g$ and the PMI type $A_g$.
To accurately distinguish the channel types, $\mathbf{X}_g$ is constructed with measurements from the current CSI-RS recived by UE $g$ that depicts channel's key features as the angle power spectrum (APS), the power ratio of the first two ranks for spatial correlation, the delay power spectrum (DPS) for frequency-selectivity and the bitmap of Type II to reflect channel sparsity in space and frequency.
Additionally, $\mathbf{X}_g$ incorporates velocity to depict user’s mobility, as well as SNR and channel quality indication (CQI) for Type I and EType II codebook, evaluating the expected performance of different codebooks. 
Based on the calculated utility in Eq.\eqref{eq:11}, the optimal codebook type $A_g$ is set as the label of $\mathbf{X}_g \in \mathcal{X}_g$.
Then, UE $g$ trains the local model $\Omega_g$ as
\begin{align} \label{eq:training}
    \mathbf{W}^{out}_g = \mathbf{W}^{out}_g - lr*\frac{\partial L}{\partial \mathbf{W}^{out}_g},
\end{align}
and sends it to the server, where
$\mathbf{W}^{out}_g$ is the output layer's weight matrix of the local model $\Omega_g$, $lr$ is the learning rate and $L$ is the loss function.
After that, the sever updates the global model $\Omega$ with $\Omega_g, g\in \mathcal{G}$ following the FedAvg as
\begin{align} \label{eq:fl}
    \Omega = \sum\nolimits_{g=1}^G \frac{|\mathcal{X}_g|}{\sum_{g=1}^G|\mathcal{X}_g|} \Omega_g,
\end{align}
and distributes $\Omega$ to each UE, as illustrated in Algorithm \ref{algorithm:1}, where
$|\mathcal{X}_g|$ is the size of the local dataset $\mathcal{X}_g$ and $\sum_{g=1}^G|\mathcal{X}_g|$ denotes the total amount of data on all UEs.

In the execution stage, UE $g$ inputs $\mathbf{X}_g$ into the trained $\Omega_g$ and outputs $A_g$ to maximize the utility in Eq. \eqref{eq:11}.
Finally, UE $g$ feeds back $A_g$ to BS to form the PDSCH precoder adaptively for each UE.
This process goes iteratively to enhance the global model $\mathbf{\Omega}$ and the adaptive striving to maximize the utility and optimize the communication performance across various channel conditions.

\begin{algorithm}[htpb]
\SetAlgoLined
    \caption{Codebook adaptation via  \systemname{}}\label{algorithm:1}
     \KwIn{Initialize the $\mathbf{\Omega}$ and $\mathbf{\Omega}_g,\forall g \in \mathcal{G}$.}

    \textit{\#\# Training Stage}
    
    \For{\rm{Each UE} $g \in \mathcal{G}$ \rm{\textbf{in parallel}}}{
    UE $g$ obtains the model from server $\mathbf{\Omega} \to \mathbf{\Omega}_g$\;
    UE $g$ constructs $\mathbf{X}_g$ and $A_g$ via Eq. \eqref{eq:11}-\eqref{eq:12} to build the training dataset\;
    UE $g$ trains $\mathbf{\Omega}_g$ with local training dataset as Eq. \eqref{eq:training}\;
    UE $g$ sends $\mathbf{\Omega}_g$ to the server\;
    Server updates the $\mathbf{\Omega}$ as Eq. \eqref{eq:fl}\;
    Server distributes $\mathbf{\Omega}$ to UEs.
  }  
  
     \textit{\#\# Execution Stage}
     
     \hspace{4mm} BS emits CSI-RS to UE $g$, $g\in \mathcal{G}$\;
     \hspace{4mm} UE $g$ constructs $\mathbf{X}_g$ into the $\mathbf{\Omega}_g$ and outputs $A_g$\;
     \hspace{4mm} UE $g$ sends $A_g$ to BS\;
     \hspace{4mm} BS forms the PDSCH precoder based on $A_g$ of the server\;

\KwOut{Adaptive codebook $A_g$ from UE$g$ for maximizing utility.}

\end{algorithm}

\begin{figure*}[ht]
\centering
\subfigure[]{
\begin{minipage}[t]{0.22\linewidth}
\includegraphics[width=1.8in]{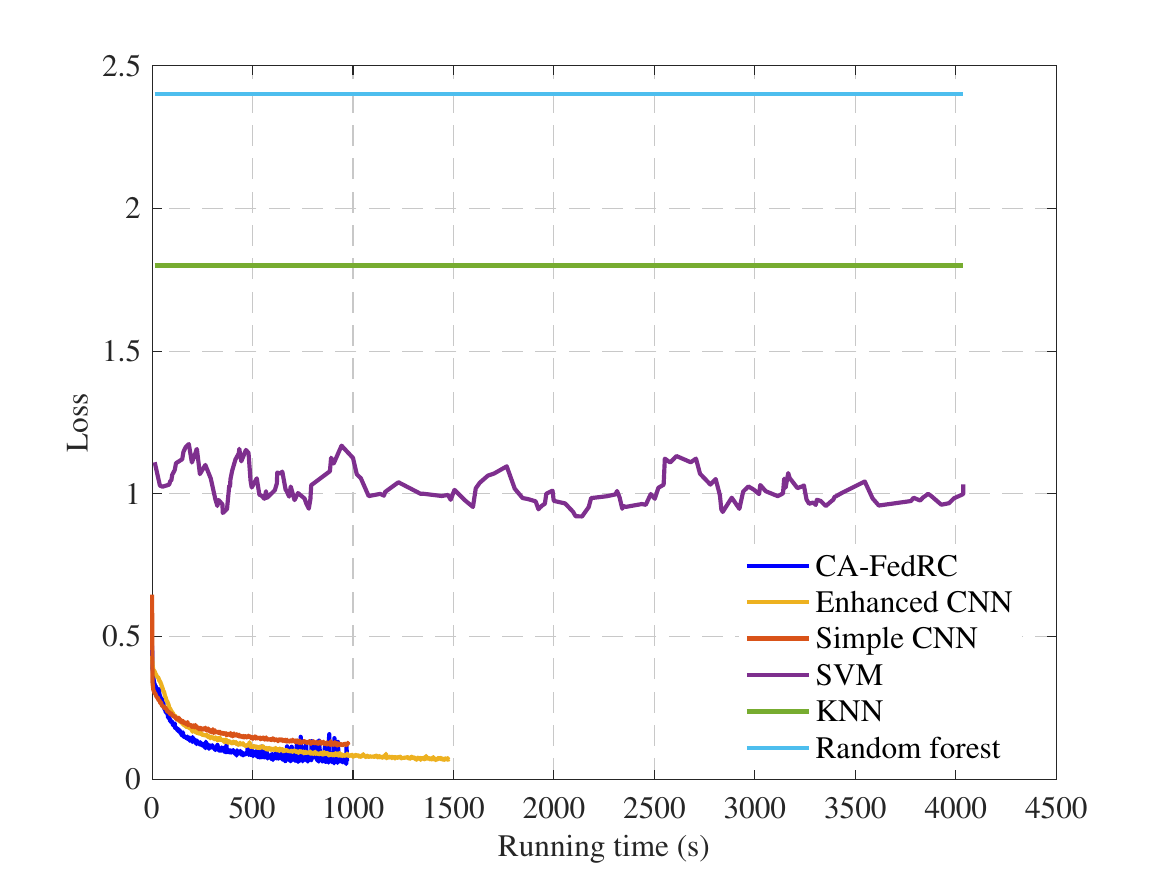}
\end{minipage}
\label{fig:2-1}
}
\subfigure[]{
\begin{minipage}[t]{0.22\linewidth}
\includegraphics[width=1.8in]{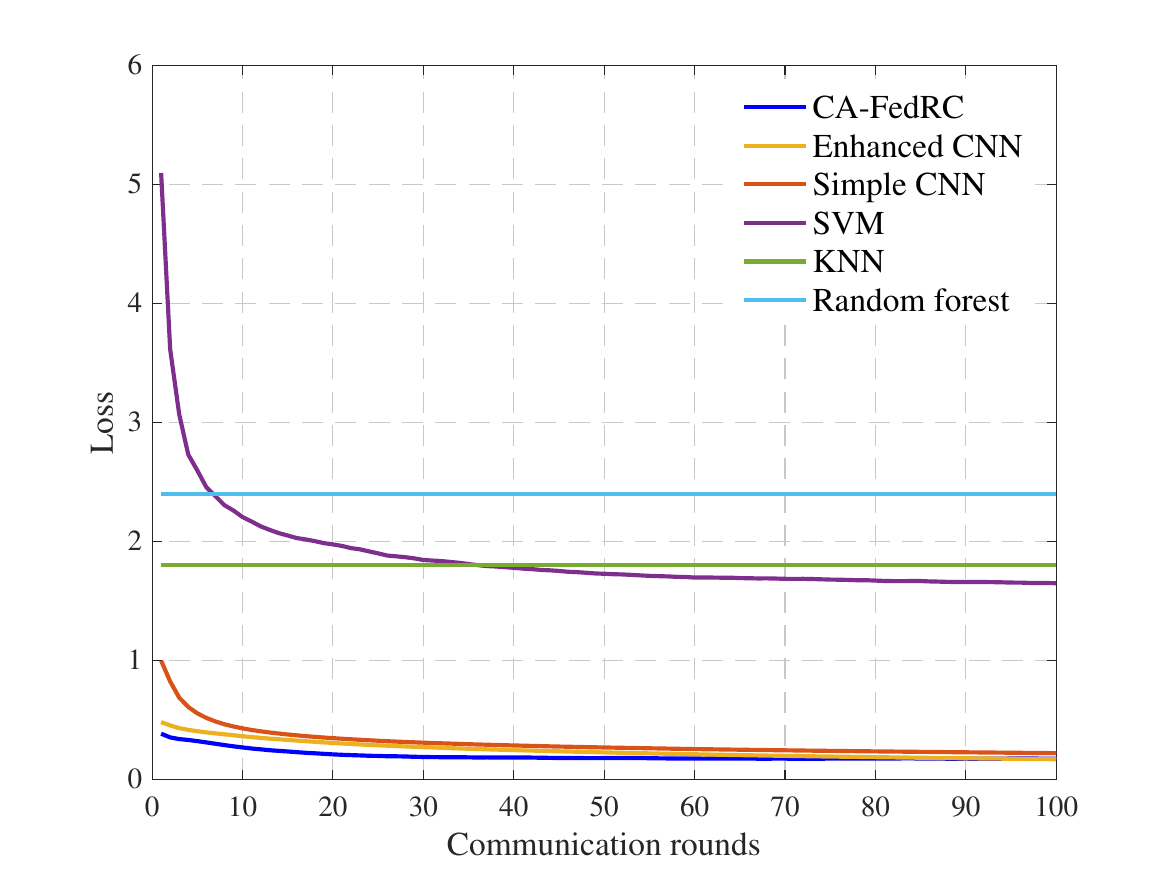}
\end{minipage}
\label{fig:2-2}
}
\subfigure[]{
\begin{minipage}[t]{0.22\linewidth}
\includegraphics[width=1.8in]{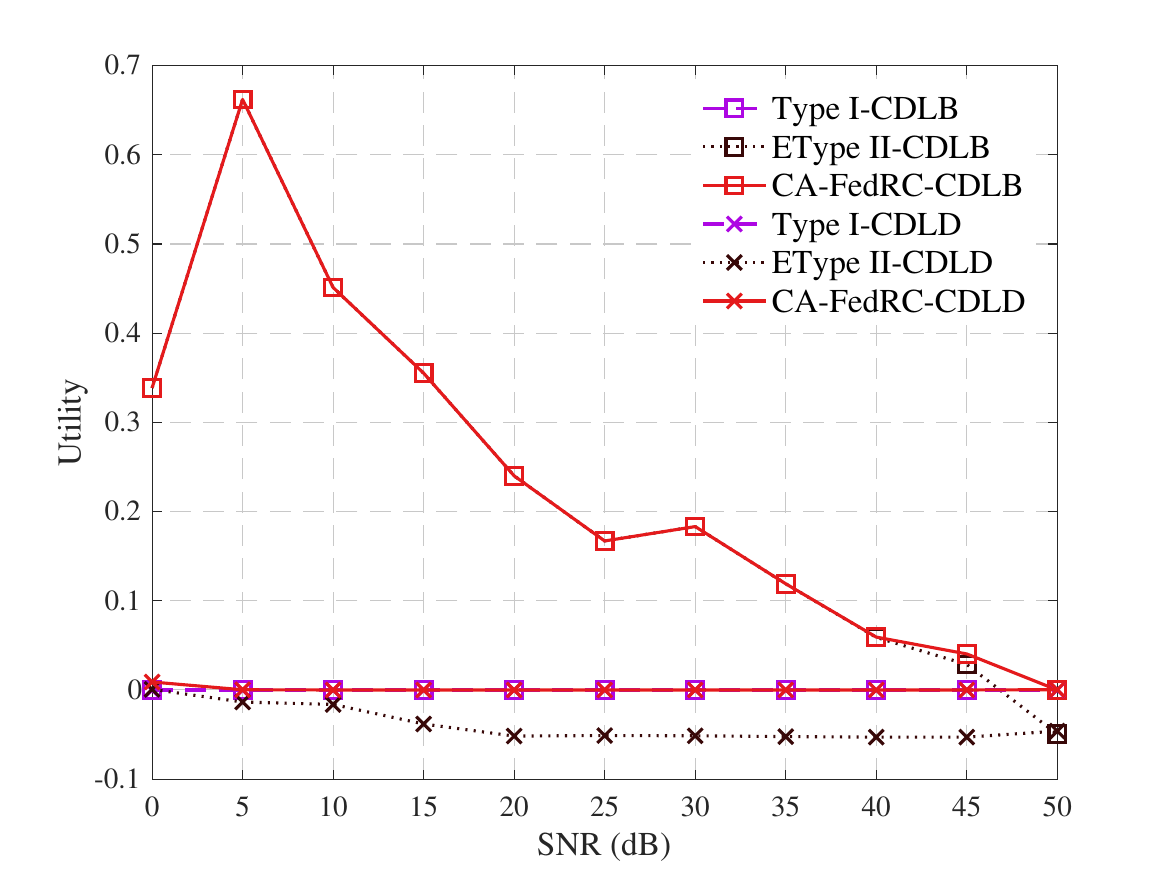}
\end{minipage}
\label{fig:2-3}
}
\subfigure[]{
\begin{minipage}[t]{0.22\linewidth}
\includegraphics[width=1.8in]{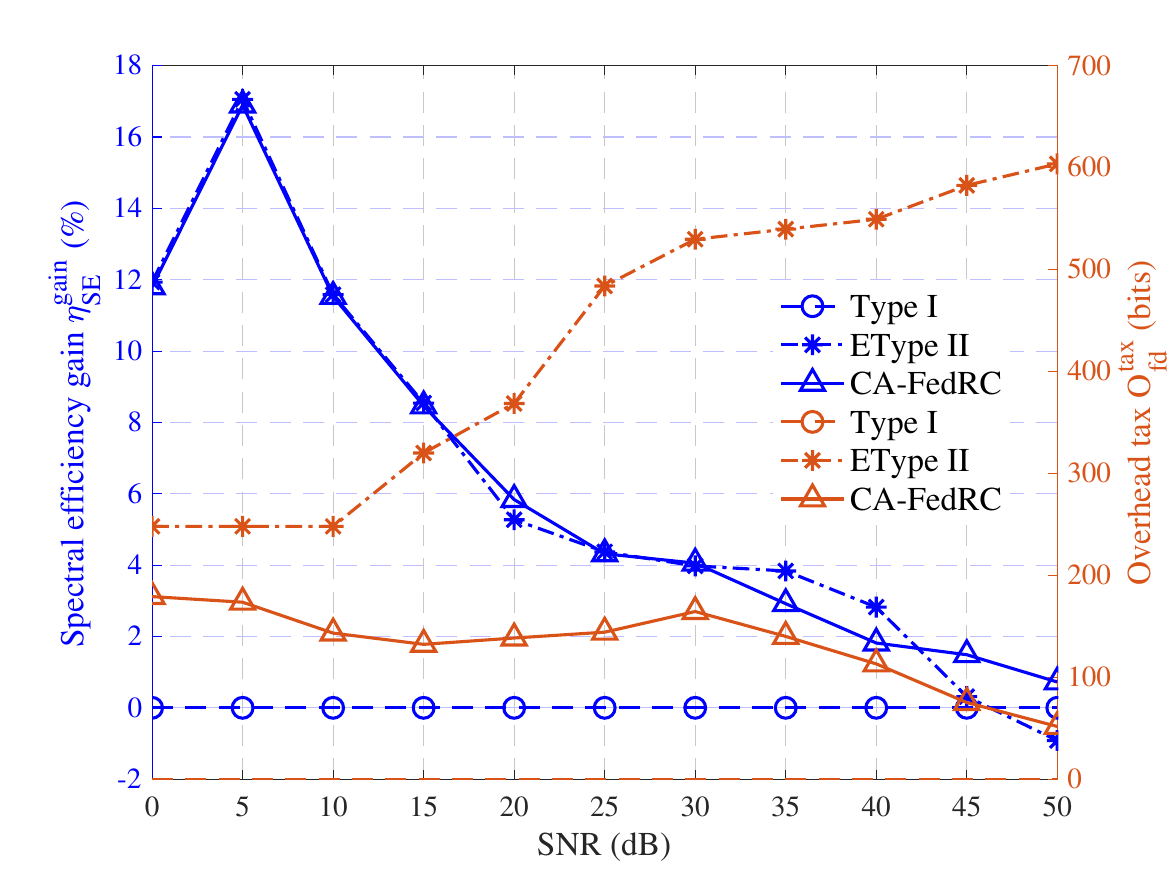}
\end{minipage}
\label{fig:2-4}
}

\caption{(a) Model comparison, (b) Federated learning process, (c) Utility for LoS and NLoS channel, and (d) Trade-off performance of \systemname{} under various channel conditions}
\vspace{-5mm}
\end{figure*}

\vspace{-7mm}
\section{Simulation Results}
We consider frequency division duplex (FDD) systems, with $N_{bs}=32$ and $N_{ue}=4$, where the uplink and downlink carrier frequency is $1.93$ GHz and $2.12$ GHz, respectively. The uplink and downlink bandwidth occupies $106$ RBs symmetrically, 
and the number of sub-band $N_{SB}=14$.
Meanwhile, $O_{fd}^{tax}$ is nearly $700$ bits when
$N_p=8$, $N_1=2$ and $N_2=8$ with $O_1=4$ and $O_2=4$.
We set $L=4 $ and $M=7$ if $r\in \{1,2\}$ or $M=4$ as $r\in \{3,4\}$ \cite{38.214}.
To balance the overhead tax and spectrum efficiency properly, we set $\lambda = 1 \times 10^{-4}$.
Each UE is equipped with the minimum mean square error-interference rejection combining (MMSE-IRC) equalizer.


The batch size for model training is $32$, learning rate is $0.001$, the default local epoch is $20$ and the number of default communication rounds for federated learning is $15$.
To examine the effectiveness and efficiency for the RC, we employ a Simple CNN as well as an Enhanced one to compare with the RC model.
Specifically, the Simple CNN is a straightforward CNN for processing sequential data which consist of two convolutional layers and one max pooling layer.
The Enhanced CNN is a more complex CNN for advanced feature extraction from sequential data which consist of three convolutional layers, each followed by batch normalization for improving stability and performance.
The neighbor number of KNN is set to $5$ and the estimator number in RF algorithm is set to $100$.
The crosstropy is set as the loss function.

In Fig. \ref{fig:2-1} and Fig. \ref{fig:2-2}, it is evident that the Reservoir Computing model, despite starting with a higher initial loss, rapidly diminishes its loss rate, outshining both the Simple and Enhanced CNN models.
It swiftly stabilizes at the lowest loss, hinting at an expedient convergence to an optimal state, a stark contrast to the Enhanced CNN which, although starting strong with the lowest initial loss, demonstrates the most gradual decline without significant improvements over time. 
Meanwhile, it remarkably outperforms SVM, KNN and RF.
Particularly in a FL setting, Reservoir Computing impressively achieves rapid convergence to low loss, outperforming the Simple CNN which converges more slowly.
Ultimately, the Reservoir Computing model converges nearly $33\%$ faster than the Enhanced CNN, showcasing not just superior predictive accuracy but also marked efficiency and suitability for energy-limited equipments.
Its swift and efficient learning curve in the FL framework, akin to the Enhanced CNN's performance, posits Reservoir Computing as a robust and resource-efficient alternative for federated models in computationally constrained environments.
\setlength{\parskip}{0pt}

As illustrated in Fig. \ref{fig:2-3}, the utility is plotted against the SNR to indicate algorithm's ability of distinguishing two typical CDLB and CDLD channels, indicating NLoS and LoS features \cite{38.901}.
For the CDLB channel type, in the high SNR region, the performance suffers significantly due to the substantial loss from compression for high rank, leading to a result that the cost does not proportional to the performance.
As a result, under these conditions, the utility for EType II is lower than Type I.
In the CDLD channel scenario, characterized by high spatial correlation, $\eta_{S\!E}^{gain}$ is negligible.
However, due to the higher overhead of EType II, the utility of EType II in the proposed model is shown to be less than that of Type I.
Nevertheless, for both channel types, the utility associated with the proposed algorithm consistently achieves the highest values, indicating the algorithm’s precise ability to recognize and differentiate between the two channel types.
\setlength{\parskip}{0pt}

In Fig. \ref{fig:2-4}, the graph portrays the practical trade-off between spectral efficiency and the feedback overhead with the setting scenarios of CDLB, CDLC and CDLD, under two different speeds at $3$ km/h, $60$ km/h and two normalization delay settings at $363$ ns and $8000$ ns, where the different curves are averaged over $90$ users and different scenarios. 
The proposed \systemname{} considers traditional Type I and EtypeII codebooks without adaptation.
When the SNR is low, our algorithm adaptively selects EType II as it significantly outperforms the baseline.
With the SNR increasing, the spectrum efficiency gain of EType II $\eta_{S\!E}^{gain}(A_g=1)$ diminishes,
while the overhead tax of adaptive codebook decreases as well.
It is observed that the proposed \systemname{} shows a spectral efficiency gain improvement of $1.6\%$ over EType II and $10\%$ over Type I averaged within $\text{SNR}\in [0,50]$ dB, while the overhead is reduced by $3.5$ times compared to EType II.
\vspace{-0.3cm}

\section{Conclusion}
We proposed a novel federated reservoir computing framework for efficient 5G NR codebook adaptation, which significantly enhances communication performance under diverse channel conditions while reducing the feedback overhead.
Meanwhile, our proposed algorithm achieves rapid convergence, significant loss reduction, and adeptly identifies channel types through innovative training data usage.
Extensive simulations confirmed its superior performance, realizing the trade-off among spectrum efficiency, computational complexity, and resource consumption, thereby advancing 5G NR efficiency and effectiveness.

\vspace{-4mm}
\bibliographystyle{ieeetr}
\bibliography{conference_101719}

\end{document}